\documentstyle[graphicx,multicol,prl,aps]{revtex}

\begin{document}
\pagestyle{empty}
\begin{multicols}{2}
\narrowtext
\parskip=0cm
\noindent
{\bf Palassini and Young reply:}
In recent work\cite{PY}, we studied the change in the ground state (GS)
configuration of a three-dimensional (3D) spin glass when the boundary
conditions (BC) are changed. Our interpretation of the data, which does
not require corrections to scaling, is that the surface of the induced domain
wall has a fractal dimension, $d_s$, less than the space dimension $d$.  This
conclusion has been convincingly demonstrated in two-dimensions (2D)
\cite{ds2D} where a much larger range of sizes can be studied.  In 3D, other
scenarios such as crossover to a behavior with $d_s = d$ on larger length
scales cannot be excluded, but these require significant corrections to
scaling.  In their comment to Ref.~\cite{PY}, Marinari and
Parisi (MP) \cite{comment}
claim that the result $d_s < d$ is improbable, and that a more likely scenario
is $d_s = d$.
However, we shall see that the additional quantities presented in \cite{comment} are affected by
strong finite-size corrections and that our original conclusion still stands:
namely $d_s<d$ is a natural interpretation because (i) it fits our data
without corrections to scaling and (ii) the same result occurs in 2D, but that
other scenarios such as $d_s=d$ cannot be ruled out.

MP consider the probability  
$P(M,L)$ that the GS configuration in an $M^{d}$ (hyper) cube is changed 
when the BC are changed from periodic to antiperiodic.
They find that in 3D
the data for $P(M,L)$ versus $M/L$ do not collapse onto a single curve,
and interpret this as evidence against $d_s<d$.
However, the 
scaling in $M/L$ is expected to hold only for $1 \ll M \ll L$, and there
are corrections both for $M \to 1$ and $M\to L$. Corrections for 
$M\to 1$ arise because the fraction of {\em bonds}
in the cube is $(M/L)^d (1- M^{-1})$, from which it follows that 
$P(M,L) \propto (M/L)^{d-d_s} (1-M^{-1})$ in this limit.
In Fig.~1a we show that in 2D
there is very good scaling for $M/L \lesssim 0.5$
when the data are rescaled by the factor $1-M^{-1}$,
but the inset shows that without this factor the data do {\em not} scale
well.
From the slope we estimate $d-d_s=0.72 \pm 0.02$ in agreement with
\cite{PY,ds2D}.
In the limit $M/L\to 1$, there is an additional correction to scaling
since the probability that the interface does {\em not} intersect 
the cube is small (presumably exponentially small in $L$). This explains
the deviation from scaling for $M/L \gtrsim 0.5$ in Fig.~1a.

In Fig.~1b we show that the 3D data is fairly similar given the
much smaller range of sizes. 
With the rescaling factor the 
data almost scale for $M/L \lesssim 0.4$.
The corrections for $M/L \gtrsim 0.4$ are larger than in 
2D, but this is reasonable since $d-d_s$ is smaller and hence $P(M,L)$
is closer to one (compare the insets in the figure).

MP also consider the fraction $P(L)$ of planes, parallel to the plane in which
the BC are changed, which are {\em not}\/ intersected by the domain wall.
$P(L)\to a$ (a constant) for large $L$ if $d_s < d$, while $P(L) \to 0$ if
$d_s=d$. MP show that $a$ is zero or very close to zero in 3D. We computed
$P(L)$ in 2D with $4\le L \le 50$, and found that $P(L)=a+b/L^c$ fits well
the data with $a=0.44 \pm 0.01$. Corrections to scaling are strong,
the deviation from the asymptotic value being $43\%$ for $L=4$, so we expect
that there are also strong corrections in 3D.  Furthermore, the value of $a$
is expected to be lower in 3D than in 2D since $d-d_s$ is smaller and, given
our lack of knowledge of $P(L)$, may actually be zero even if $d_s < d$.

We thank E. Marinari and G. Parisi for a useful correspondence and for 
sending us the data in Fig.~1b.

\begin{figure}
\centering
\includegraphics[width=0.33\textwidth]{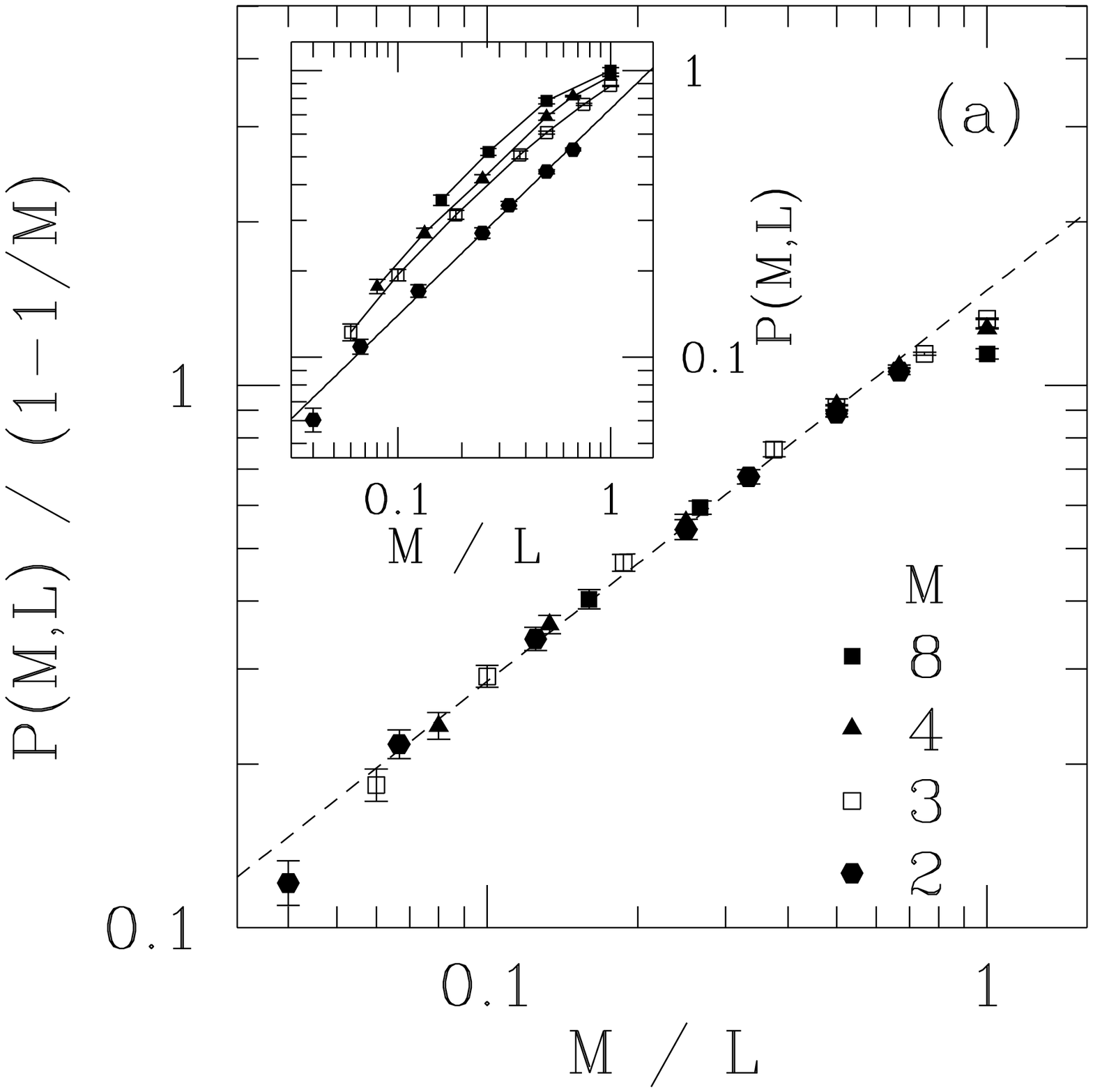}
\includegraphics[width=0.33\textwidth]{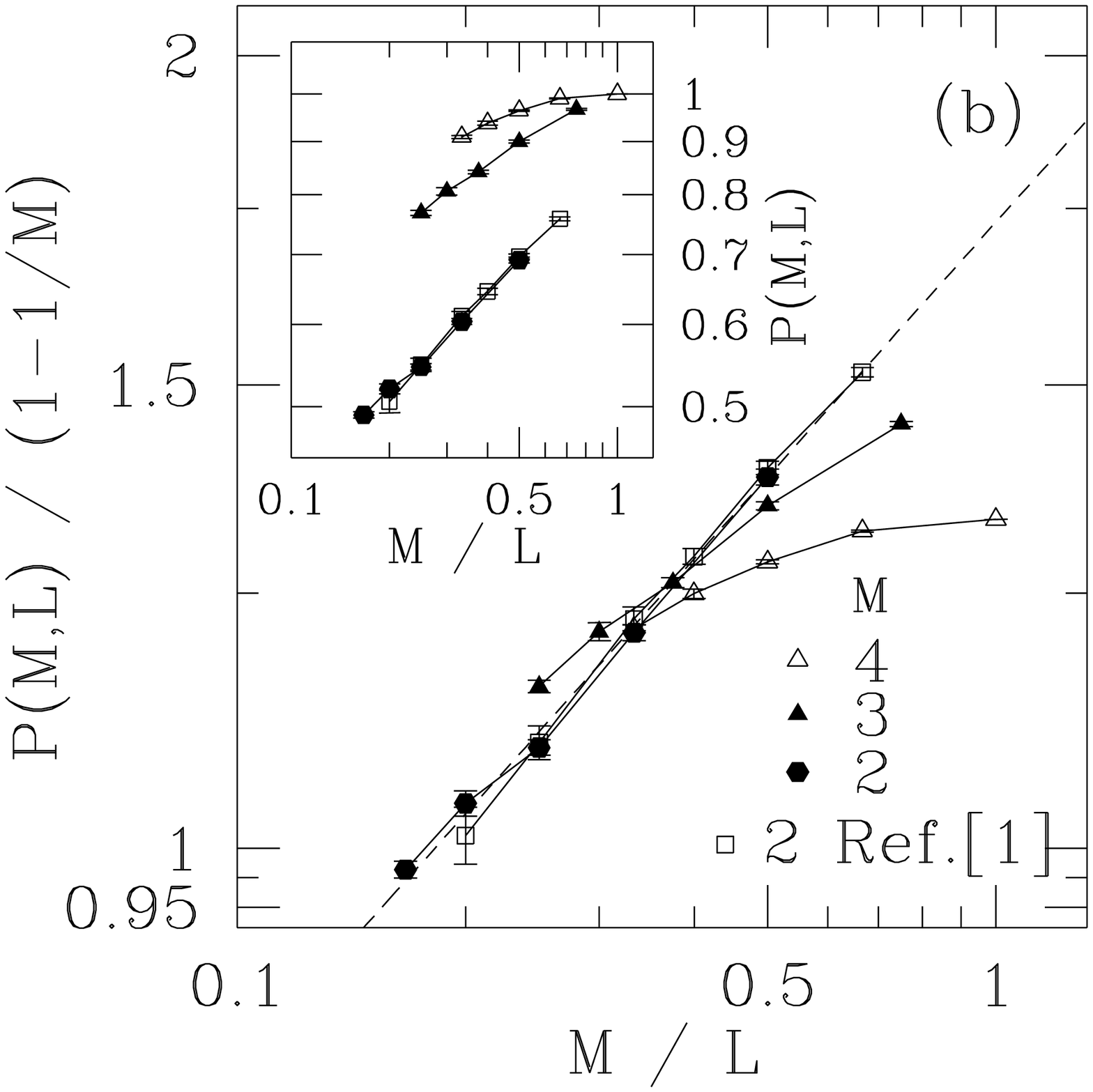}
\caption{(a) Scaling plot of $P(M,L)$ versus $ML^{-1}$ in 2D for $3 \le L \le
50$.
The dashed line has slope 0.72. Inset: scaling plot without
the rescaling factor $1-M^{-1}$: (b) The same but for 3D and using the data of
\protect\cite{comment} for $4\le L\le 12$. The dashed line has slope 0.32.
}
\end{figure}

\noindent 
Matteo Palassini and A.P. Young, \\
\indent {\small   University of California Santa Cruz}

\end{multicols}
\end{document}